\documentclass[a4paper]{article}
\usepackage{amsmath,amssymb,amsthm}
\usepackage{graphicx}
\usepackage{algorithm,algorithmic}
\usepackage{setspace,enumerate}
\usepackage{authblk}

\newtheorem{lemma}{Lemma}

\newcommand{\figurefolder}{figures}
\newcommand{\expfolder}{exp}
\newcommand{\figurescale}{0.6}
\newcommand{\expscale}{0.45}

\usepackage{setspace,soul}
\usepackage[dvipsnames]{xcolor}
\sethlcolor{Goldenrod}
\soulregister{\cite}{1}
\soulregister{\ref}{1}

\newcommand{\R}{\ensuremath{\mathcal{R}}}
\newcommand{\D}{\ensuremath{\mathcal{D}}}
\newcommand{\g}{\ensuremath{\mathcal{G}}}
\newcommand{\f}{\ensuremath{\phi}}

\renewcommand{\^}[1]{\hat{#1}}

\newcommand{\capfontsize}{\normalsize} %{\small}
\newcommand{\tablefontsize}{\normalsize} %{\small}
\DeclareMathOperator*{\argmin}{argmin}

\title{Efficient Exact $k$-Flexible Aggregate Nearest Neighbor Search in Road Networks Using the M-tree}
\author[1,2]{Moonyoung Chung} \author[1]{Soon J. Hyun} \author[3]{Woong-Kee Loh}
\affil[1]{\normalsize School of Computing, Korea Advanced Institute of Science and Technology (KAIST), Daejeon 34141, Republic of Korea}
\affil[2]{\normalsize Artificial Intelligence Research Laboratory, Electronics and Telecommunications Research Institute (ETRI), Daejeon 34129, Republic of Korea}
\affil[3]{\normalsize School of Computing, Gachon University, Seongnam 13120, Republic of Korea }
\begin{document}
\maketitle
\doublespacing

\begin{abstract}
This study proposes an efficient exact $k$-flexible aggregate nearest neighbor ($k$-FANN) search algorithm in road networks using the M-tree. The state-of-the-art IER-$k$NN algorithm used the R-tree and pruned off unnecessary nodes based on the Euclidean coordinates of objects in road networks. However, IER-$k$NN made many unnecessary accesses to index nodes since the Euclidean distances between objects are significantly different from the actual shortest-path distances between them.
In contrast, our algorithm proposed in this study can greatly reduce unnecessary accesses to index nodes compared with IER-$k$NN since the M-tree is constructed based on the actual shortest-path distances between objects. To the best of our knowledge, our algorithm is the first exact FANN algorithm that uses the M-tree. We prove that our algorithm does not cause any false drop. In conducting a series of experiments using various real road network datasets, our algorithm consistently outperformed IER-$k$NN by up to 6.92 times.
%\begin{keywords}
%\end{keywords}
\end{abstract}

\section{Introduction}
\label{sec:introduction}

This study proposes an efficient $k$-\textit{flexible aggregate nearest neighbor} (\textit{FANN}) search algorithm~$(k \geq 1)$. The FANN search is an extension of the \textit{aggregate nearest neighbor} (\textit{ANN}) search, which is also an extension of the traditional \textit{nearest neighbor} (\textit{NN}) search. The NN search, which finds the object closest to the given query object $q$ among the objects in a dataset $\D$, is an important subject pursued in many applications in various domains~\cite{abe16, kri07, sha07}. The ANN search~\cite{iou07, mia20, yiu05}, which extends the NN search by introducing a query set $Q$ including $M ~ (\geq 1)$ query objects $q_j ~ (0 \leq j < M)$, finds an object $p^*$ that satisfies the following Eq.~(\ref{eqn01}):
\begin{eqnarray}
p^* = \argmin_{p_i \in \D} \left \{ \g \left \{ d(p_i, q_j), q_j \in Q \right \} \right \}, \label{eqn01}
\end{eqnarray}
where $\g$ denotes an aggregate function such as max and sum, and $d()$ denotes the distance between two objects. An example of applying ANN search is to find an optimal place for a meeting of $M$ members.

The FANN search~\cite{li11, li16, yao18}, which extends ANN search by introducing a flexibility factor $\phi ~ (0 < \phi \leq 1)$, finds an object $p^*$ that satisfies the following Eq.~(\ref{eqn02}):
\begin{eqnarray}
p^* = \argmin_{p_i \in \D} \left \{ \g \left \{ d(p_i, q_j), q_j \in Q_\phi \right \} \right \}, \label{eqn02}
\end{eqnarray}
where $Q_\phi$ denotes any subset of $Q$ of $\phi M$ size. An example of FANN search is to find an optimal place for a meeting of $\phi M$ members, the minimum quorum of $M$ members. The FANN search cannot be solved simply by running an ANN search algorithm for every possible $Q_\phi$. For example, in the case of $M$ = 256 and $\phi$ = 0.5, ANN search must be performed as much as $5.769 \times 10^{75}$ times.
In this study, the target of FANN search is the objects in a points-of-interest (POIs) set $P ~ (\subseteq \D)$, e.g., hospitals and restaurants, instead of the whole dataset $\D$.

The existing ANN and FANN search algorithms have been studied separately for Euclidean spaces and road networks. A road network is represented with a graph data structure, and the distance between two objects is defined as the shortest-path distance between them~\cite{abe16, yao18, yiu05}. Since the calculation of the shortest-path distance has a much higher complexity than that of the Euclidean distance~\cite{chu21, kri08, zho15b}, ANN and FANN search algorithms in road networks should minimize the calculations of the shortest-path distances. Yao et al.~\cite{yao18} proposed a few algorithms for exact FANN search in road networks, and among them, the \textit{IER-$k$NN} algorithm showed the highest performance. It used the R-tree~\cite{man05} and pruned off the nodes that are unlikely to include the final result objects, thus reducing the calculations of the shortest-path distances for the objects in the pruned nodes. Nevertheless, when deciding whether to prune a specific node, IER-$k$NN accessed many unnecessary nodes since the decision is made using the Euclidean distances, which are significantly different from the actual shortest-path distances, and thus performs many shortest-path distance calculations for objects included in the unnecessary nodes.

This study proposes an efficient exact $k$-FANN search algorithm using the M-tree~\cite{cia97} and proves that the proposed algorithm does not cause any false drop. While the R-tree is an index structure for objects in a Euclidean space, the M-tree is constructed for a dataset in a metric space, where a distance function between objects is given instead of their actual coordinates. The road network can be mapped into a metric space~\cite{iou07, yao18}, and the M-tree is constructed using the actual shortest-path distances between objects in road networks. Therefore, our algorithm can prune the index nodes more accurately than the state-of-the-art IER-$k$NN algorithm and can dramatically reduce the calculations of the shortest-path distances. To the best of our knowledge, our algorithm is the first exact FANN algorithm that uses the M-tree.
The performance of our algorithm is compared with that of IER-$k$NN using various real road network datasets. The experimental result demonstrated that our algorithm consistently outperformed IER-$k$NN for all datasets and parameters, with a performance improvement of up to 6.92 times.

This paper is organized as follows. Section~\ref{sec:relatedWork} briefly explains the structure of the M-tree and the existing FANN search algorithms. Section~\ref{sec:proposedAlgorithm} describes our algorithm in detail. Section~\ref{sec:evaluation} compares the search performance for various real road network datasets and parameters. Finally, Section~\ref{sec:conclusions} concludes this study.

\section{Related Work}
\label{sec:relatedWork}

In this section, we discuss various previous NN, ANN, and FANN algorithms and then briefly explain the structure of the M-tree, which is necessary for describing our algorithm.
With the recent spread of ubiquitous mobile devices, the demand for efficient $k$-NN search in road networks has increased. Abeywickrama et al.~\cite{abe16} evaluated the performance of various existing $k$-NN algorithms including Incremental Network Expansion (INE)~\cite{pap03}, Incremental Euclidean Restriction (IER)~\cite{pap03}, Route Overlay and Association Directory (ROAD)~\cite{lee12}, and G-tree~\cite{zho15a}. They demonstrated as an experimental result using synthetic and real road network datasets that the best performance was achieved with the combination of the previously neglected IER algorithm and \textit{pruned highway labeling} (\textit{PHL}) algorithm~\cite{abr11, aki14}. Shaw et al.~\cite{sha07} presented an approximate $k$-NN algorithm using Road Network Embedding (RNE), which maps objects on a road network to a $p$-dimensional Euclidean space. The algorithm stored the mapped objects in the M-tree and showed the search performance superior to the existing INE~\cite{pap03} algorithm.

Gao et al.~\cite{gao15} dealt with the reverse $k$-NN (R$k$NN) problem in road networks. They presented an algorithm based on a heuristic filter-and-refinement framework that simultaneously considers spatial and textual information and demonstrated its efficiency using synthetic and real datasets. Zhao et al.~\cite{zha20} dealt with the problem of diversified top-$k$ geo-social keyword (D$k$GSK) query that considers spatial, social, and textual constraints between the query and data objects. They considered not only the relevance but also the diversity of the query result in order to enhance the quality of the result. They showed that the problem was NP-hard and proposed an exact algorithm based on several heuristics and an approximate algorithm, whose efficiency was demonstrated using actual datasets.

Li et al.~\cite{li11, li16} addressed the FANN search problem in a Euclidean space and proposed algorithms using the R-tree and a list data structure. The R-tree-based algorithm estimates the FANN distance based on the Euclidean distance between $\phi M$ query objects that are nearest to the MBR of each R-tree node and determines whether to prune the node based on the estimated distance. The list-based algorithm finds the final FANN object while gradually constructing the nearest object list for each query object $q_i$. Li et al.~\cite{li11, li16} conducted various experiments for the algorithms and showed that the R-tree-based algorithm always had a higher search performance.

Ioup et al.~\cite{iou07} proposed an ANN search algorithm in road networks using the M-tree~\cite{cia97}. However, this algorithm only returns an approximate result, and the error ratio of the search result is unknown.
Miao et al.~\cite{mia20} dealt with the continuous $k$-ANN (CA$k$NN) problem in dynamic road networks, where the locations of data and query objects and the edge weights are changing. They defined \textit{partial distance matrix} data structure that contains only data objects closer than the \textit{safe distance} $r$ from each query object, where $r$ is not greater than the aggregate distance of the $k$-th candidate ANN object. They showed that their algorithm was superior to the existing algorithm that assumes static query objects through experiments using actual road network datasets.

The FANN search problem in road networks was addressed by Yao et al.~\cite{yao18}. They proposed the Dijkstra-based algorithm, R-List algorithm, and R-tree-based IER-$k$NN algorithm. In addition, they presented an exact algorithm that does not require an index for $\g$ = max. They experimentally showed that IER-$k$NN had the best performance for all parameters and road network datasets. However, IER-$k$NN accessed many unnecessary R-tree nodes and performed many unnecessary shortest-path distance calculations for objects included in the unnecessary nodes. The algorithms that did not use an index showed a much lower search performance than the algorithms using an index.
Chen et al.~\cite{che20} addressed the FANN search problem that took keyword similarity into account in road networks. They defined a new distance function based on both the aggregate of distances to query objects $q_i ~ (\in Q_\phi)$ and keyword similarity. They presented algorithms (denoted as \textit{KFANN}) by extending the Dijkstra-based algorithm, R-List algorithm, and IER-kNN previously proposed by Yao et al.~\cite{yao18}.

\begin{figure}
\centering \capfontsize
\includegraphics[scale=\figurescale]{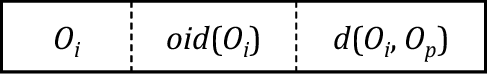} \\
(a) Leaf node entry. \\[0.1in]
\includegraphics[scale=\figurescale]{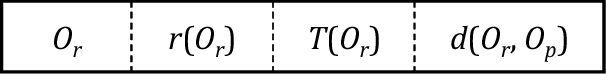} \\
(b) Non-leaf node entry.
\caption{Structures of M-tree node entries.} \label{fig21}
\end{figure}

The M-tree~\cite{cia97} is a balanced tree index structure similar to the R-tree~\cite{man05}. While the region for a node of the R-tree is a minimum bounding rectangle (MBR) including all entries in the corresponding node, the region for a node of the M-tree is a sphere defined by a central object (or parent object) and radius. Figure~\ref{fig21}(a) shows the structure of an entry in an M-tree leaf node. A leaf entry corresponds to an object in a dataset. In Figure~\ref{fig21}(a), $O_i$ is an object, $oid(O_i)$ is the object ID of $O_i$, and $d(O_i,O_p)$ is the distance between $O_i$ and the parent object $O_p$. The \textit{parent object} $O_p$ is a central object that represents a leaf node $L$; among all the objects $O_i$ in $L$, $O_p$ is chosen such that it satisfies the following Eq.~(\ref{eqn1}):
\begin{eqnarray}
O_p = \argmin_{O_i \in L} \left \{ \max \left \{ d(O_i, O_j), O_j \in L \right \} \right \}. \label{eqn1}
\end{eqnarray}

Figure~\ref{fig21}(b) shows the structure of an entry in an M-tree non-leaf node $N$. A non-leaf entry corresponds to a sub-node $n$ of $N$. In Figure~\ref{fig21}(b), $O_r$ is called the \textit{routing object} and set as the parent object of $n$. $r(O_r)$ is the radius of the spherical region of $n$, $T(O_r)$ is a pointer to the subtree rooted by $n$, and $d(O_r, O_p)$ is the distance between $O_r$ and $O_p$ of $N$. The parent object $O_p$ is chosen as the routing object $e_p.O_r$ of the entry $e_p$ such that it satisfies the following Eq.~(\ref{eqn2}) among the entries $e_i$ in $N$:
\begin{eqnarray}
e_p = \argmin_{e_i \in N} \left \{ \max \left \{ d(e_i, e_j), e_j \in N \right \} \right \}, \label{eqn2} \\
d(e_i, e_j) = d(e_i.O_r, e_j.O_r) + e_i.r(O_r) + e_j.r(O_r). \label{eqn3}
\end{eqnarray}

\section{FANN-PHL: Proposed $k$-FANN Algorithm}
\label{sec:proposedAlgorithm}

In this section, we explain our exact $k$-FANN search algorithm that uses the M-tree constructed using the actual shortest-path distance $D$ between two objects in a road network. To obtain the distance $D$ between two objects, we used the PHL algorithm~\cite{abr11, aki14}, which is known as the fastest algorithm to obtain $D$~\cite{abe16, yao18}. Our algorithm is referred to as \textit{FANN-PHL} hereafter. Table~\ref{notation} summarizes the notations used in this study.

\begin{table}
\caption{Summary of notations.} \label{notation}
\centering \tablefontsize
\begin{tabular}{cl} \hline
Notation & Description \\ \hline
$\R$ & road network dataset \\
%$N$ & number of objects (vertices) in $\R$ \\
$D$ & shortest-path distance between objects in $\R$ \\
$Q$ & set of query objects \\
$M$ & number of query objects, i.e., $M = |Q|$ \\
$\phi$ & flexibility factor $(0 < \phi \leq 1)$ \\ \hline
\end{tabular}
\end{table}

Algorithm~\ref{alg1} describes the FANN-PHL algorithm, which has an overall structure similar to that of the previous FANN algorithms~\cite{li11, li16, yao18}. The input of the algorithm consists of a road network $\R$, a POI set $P ~ (\subseteq \D)$, a query object set $Q$, flexibility factor $\phi$, aggregate function $\g$ (= max or sum), and an M-tree $T$. The algorithm returns the FANN object $p^*$, a query subset $Q^*_\f$, and the FANN distance $g(p^*, Q^*_\f)$, where $g(p^*, Q^*_\f) = \g \{ d(p^*, q_j), q_j \in Q^*_\f \}$. Algorithm~\ref{alg1} is for the case in which the number of FANN objects $k$ is 1, and the natural extension for the general case of $k \geq 1$ will be described later in this section.

\begin{algorithm}[t]
\begin{algorithmic}[1]
\REQUIRE $\R, P, Q, \f, \g, T$
\ENSURE $p^*, Q^*_\f, g(p^*, Q^*_\f)$
\STATE $\^{p}^*.g_\f \leftarrow \infty, H \leftarrow \varnothing$ \label{l1.01}
\STATE $H.push(e)$ for all entries $e$ in $T.root$ \label{l1.02}
\WHILE{$H \neq \varnothing$} \label{l1.03}
\STATE $e \leftarrow H.pop()$ \label{l1.04}
\IF{$e.n$ is a non-leaf node} \label{l1.05}
\FOR{each entry $e'$ in $e.n$}
\IF{$e'.G_\f \leq \^{p}^*.g_\f$} \label{l1.07}
\STATE \textbf{if} $e'.g_\f \leq \^{p}^*.g_\f$ \textbf{then} $H.push(e')$ \textbf{end if} \label{l1.08}
\ENDIF
\ENDFOR
\ELSE
\FOR{each object $p$ in $e.n$ such that $p \in P$}
\IF{$p.G_\f \leq \^{p}^*.g_\f$} \label{l1.13}
\STATE \textbf{if} $p.g_\f \leq \^{p}^*.g_\f$ \textbf{then} $\^{p}^* \leftarrow p$ \textbf{end if} \label{l1.14}
\ENDIF
\ENDFOR
\ENDIF
\ENDWHILE \label{l1.19}
\STATE Return $\^{p}^*$ \label{l1.20}
\end{algorithmic}
\caption{FANN-PHL Algorithm.}
\label{alg1}
\end{algorithm}

We explain each line of Algorithm~\ref{alg1} in detail. In line~\ref{l1.01}, $\^{p}^*$ denotes the FANN object that has been found until now during the execution of FANN-PHL, and its FANN distance $\hat{p}^*.g_\phi$ is initialized as $\infty$. $H$ is a priority queue that includes the M-tree non-leaf node entries. In line~\ref{l1.02}, all entries of the root node of the M-tree are inserted into $H$.
The while loop in lines~\ref{l1.03} $\sim$ \ref{l1.19} is repeated until there is no entry remaining in $H$. In line~\ref{l1.04}, the entry that has the highest priority in $H$, i.e., the entry that has the highest possibility of including the final FANN object is extracted. Here, the possibility for a specific entry $e$ is estimated using its FANN distance $e.g_\f$, and the entry with the smallest $e.g_\f$ distance among the entries in $H$ is extracted. The $e.g_\f$ distance can be obtained using Eq.~(\ref{eqn31}) below.

\begin{figure}
\centering \capfontsize
\includegraphics[scale=\figurescale]{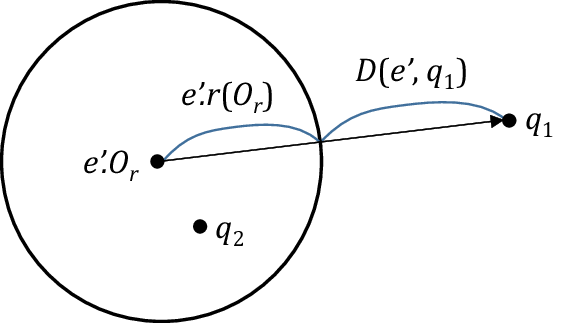}
\caption{Distances $D$ between an entry $e'$ and query objects $q_1$ and $q_2$.} \label{fig32}
\end{figure}

In line~\ref{l1.05}, $e.n$ is the sub-node for entry $e$, i.e., the root node of the subtree pointed by $e.T(O_r)$ in Figure~\ref{fig21}(b). If the node $e.n$ is a non-leaf node, in line~\ref{l1.08}, the possibility of including the final FANN object is estimated for each entry $e'$ in $e.n$; if there exists any possibility, $e'$ is inserted into $H$. To estimate the possibility, the FANN distance $e'.g_\f$ for each entry $e'$ is calculated as follows:
\begin{eqnarray}
e'.g_\f = \min \left \{ g(e', Q_\f), Q_\f \subseteq Q \right \}, \label{eqn31} \\
g(e', Q_\f) = \g \{ D(e', q_i), q_i \in Q_\f \}, \label{eqn32}
\end{eqnarray}
where $D(e', q_i)$ is the distance between the spherical region for the node $e'.n$ and a query object $q_i$, and is defined as $D(e', q_i) = \max \{ D(e'.O_r, q_i) - e'.r(O_r), 0 \}$. Figure~\ref{fig32} shows $D(e', q_i)$ for two query objects $q_1$ and $q_2$. The FANN distance of an object included within the spherical region such as $q_2$ is defined as zero.

\begin{figure}
\centering \capfontsize
\includegraphics[scale=\figurescale]{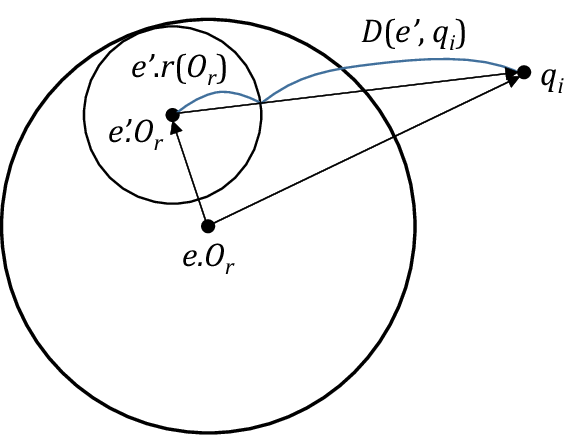} \\
(a) In a non-leaf node. \\[0.1in]
\includegraphics[scale=\figurescale]{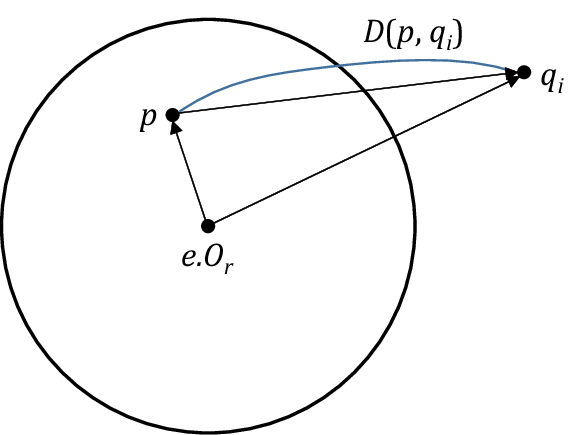} \\
(b) In a leaf node.
\caption{Finding entries/objects to prune in FANN-PHL.} \label{fig33}
\end{figure}

In line~\ref{l1.08}, if the FANN distance $e'.g_\f$ for a specific entry $e'$ is smaller than the FANN distance $\^{p}^*.g_\f$ of the FANN candidate object $\^{p}^*$ that has been found until now, $e'$ is inserted into $H$ together with $e'.g_\phi$. The $D(e'.O_r, q_i)$ distance required to obtain $e'.g_\f$ is the shortest-path distance between two objects $e'.O_r$ and $q_i$, and its calculation is expensive as explained above. Hence, in line~\ref{l1.07}, the entries without the possibility of including the final FANN object are pruned off at a lower cost. For each entry $e'$, $e'.G_\f$ is calculated as follows:
\begin{eqnarray}
e'.G_\f = \min \left \{ G(e', Q_\f), Q_\f \subseteq Q \right \}, \label{eqn33} \\
G(e', Q_\f) = \g \{ D_G(e', q_i), q_i \in Q_\f \}, \label{eqn34}
\end{eqnarray}
where $D_G(e', q_i)$ is the distance between the spherical region for a node $e'.n$ and a query object $q_i$ and is defined as $D_G(e', q_i) = | D(e.O_r, q_i) - D(e.O_r, e'.O_r) | - e'.r(O_r)$ (see Figure~\ref{fig33}(a)). The only difference from Eqs.~(\ref{eqn31}) and (\ref{eqn32}) is that $D$ is used in Eqs.~(\ref{eqn31}) and (\ref{eqn32}) whereas $D_G$ is used in Eqs.~(\ref{eqn33}) and (\ref{eqn34}).
Since $e.O_r$ is the parent object in node $n$, which includes $e'$, $D(e.O_r, e'.O_r) = D(e'.O_p, e'.O_r)$ and is already stored in $e'$ together with $e'.r(O_r)$ (see Figure~\ref{fig21}(b)). The distance $D(e.O_r, q_i)$ can be used commonly for every $e'$ once it is calculated; therefore, it can reduce the calculations of $D$ distances.

In line~\ref{l1.05}, if $e.n$ is a leaf node, in line~\ref{l1.14}, the FANN distance $p.g_\f$ is calculated as follows for each object $p$ in $e.n$:
\begin{eqnarray}
p.g_\f = \min \left \{ g(p, Q_\f), Q_\f \subseteq Q \right \}, \label{eqn11} \\
g(p, Q_\f) = \g \{ D(p, q_i), q_i \in Q_\f \}. \label{eqn12}
\end{eqnarray}
Here, it should be checked if the object $p$ belongs to the POI set $P$. If the FANN distance of $p$ is smaller than that of the FANN candidate object $\^{p}^*$, $p$ is set as a new FANN candidate object.
The cost of calculating the FANN distance of an object $p$ is very high since the distance $D$ between $p$ and all query objects $q_i$ should be obtained. Hence, in line~\ref{l1.13}, as in line~\ref{l1.07}, the objects that are unlikely to be FANN objects are pruned off at a lower cost. That is, $p.G_\f$ is calculated for each object $p$ as follows:
\begin{eqnarray}
p.G_\f = \min \left \{ G(p, Q_\f), Q_\f \subseteq Q \right \}, \label{eqn35} \\
G(p, Q_\f) = \g \{ D_G(p, q_i), q_i \in Q_\f \}, \label{eqn36}
\end{eqnarray}
where $D_G(p, q_i) = | D(e.O_r, q_i) - D(e.O_r, p) |$ (see Figure~\ref{fig33}(b)). The only difference from Eqs.~(\ref{eqn11}) and (\ref{eqn12}) is that $D$ is used in Eqs.~(\ref{eqn11}) and (\ref{eqn12}) whereas $D_G$ is used in Eqs.~(\ref{eqn35}) and (\ref{eqn36}).
Since $e.O_r$ is the parent object in node $n$, which includes $p$, $D(e.O_r, p) = D(O_p, p)$ and is already stored in the leaf node entry for $p$ (see Figure~\ref{fig21}(a)). The calculations of $D$ distances can be reduced since $D(e.O_r, q_i)$ is commonly used for every $p$ once it is calculated.
In line~\ref{l1.20}, the FANN candidate object $\^{p}^*$ is returned as the final FANN object $p^*$. The following Lemma~\ref{lem1} proves that the FANN-PHL algorithm is correct.

\begin{lemma} \label{lem1}
The FANN-PHL algorithm does not cause any false drop.
\begin{proof}
In line~\ref{l1.08}, since it holds that $D(p, q_i) \geq D(e', q_i) ~ (0\leq i<M)$ for any object $p$ included in the spherical region for $e'$, it holds that $g(p, Q_\f) \geq g(e', Q_\f)$, i.e., $p.g_\f \geq e'.g_\f$ for any $Q_\f$ (see Figure~\ref{fig33}(a)). If the condition in line~\ref{l1.08} is not satisfied for the FANN candidate object $\^{p}^*$, i.e., if $e'.g_\f > \^{p}^*.g_\f$, it holds that $p.g_\f > \^{p}^*.g_\f$ for any object $p$ in $e'$. Therefore, $e'$ can be safely discarded.

In line~\ref{l1.07}, it is always true that $D(e', q_i) + e'.r(O_r) \geq | D(e.O_r, q_i) - D(e.O_r, e'.O_r) |$, i.e., $D(e', q_i) \geq | D(e.O_r, q_i) - D(e.O_r, e'.O_r) | - e'.r(O_r) = D_G(e', q_i) ~ (0\leq i<M)$ (see Figure~\ref{fig33}(a)). Hence, it holds that $g(e', Q_\f) \geq G(e', Q_\f)$, i.e., $e'.g_\f \geq e'.G_\f$ for any $Q_\f$. If the condition in line~\ref{l1.07} is not satisfied, i.e., if $e'.G_\f > \^{p}^*.g_\f$, it holds that $e'.g_\f > \^{p}^*.g_\f$. Therefore, $e'$ can be discarded safely based on the proof for line~\ref{l1.08}.

In line~\ref{l1.13}, it is always true that $D(p, q_i) \geq | D(e.O_r, q_i) - D(e.O_r, p) | = D_G(p, q_i) ~ (0\leq i<M)$ (see Figure~\ref{fig33}(b)). Hence, for any $Q_\f$, it holds that $g(p, Q_\f) \geq G(p, Q_\f)$, i.e., $p.g_\f \geq p.G_\f$. If the condition in line~\ref{l1.13} is not satisfied, i.e., if $p.G_\f > \^{p}^*.g_\f$, it holds that $p.g_\f > \^{p}^*.g_\f$, and therefore $p$ can also be discarded safely.

In conclusion, considering all the aforementioned proofs together, the FANN-PHL algorithm in Algorithm~\ref{alg1} does not cause any false drop.
\end{proof}
\end{lemma}

Algorithm~\ref{alg1} applies to the case where the number of FANN objects $k$ is 1, and it can be extended to the general case of $k \geq 1$ as follows. First, an array $K$ is allocated to store $k$ FANN result objects and initialized as $K_i.g_\f = \infty ~ (0 \leq i < k)$. The FANN candidate objects in $K$ are always ordered by their respective $K_i.g_\f$ values. In lines~\ref{l1.07}, \ref{l1.08}, \ref{l1.13}, and \ref{l1.14} in Algorithm~\ref{alg1}, comparisons are made with $K_{k-1}.g_\f$ instead of $\^{p}^*.g_\f$. When the condition in line~\ref{l1.14} is satisfied, a new object $p$ is inserted into $K$, and the previous object $K_{k-1}$ is removed. Finally, the array $K$ is returned in line~\ref{l1.20}.

\section{Experimental Evaluation}
\label{sec:evaluation}

In this section, we compare the search performance of our FANN-PHL algorithm with that of the IER-$k$NN algorithm~\cite{yao18} through a series of experiments using real road network datasets. The platform is a workstation with AMD 3970X CPU, 128GB memory, and 1.2TB SSD. We implemented both FANN-PHL and IER-$k$NN in C/C++.

The datasets used in the experiments are real road networks of five regions in the U.S. These datasets have been used in the 9th DIMACS Implementation Challenge $-$ Shortest Paths\footnote{http://www.diag.uniroma1.it/challenge9/download.shtml} and many previous studies~\cite{abe16, yao18}. Table~\ref{tab41} summarizes the datasets used in the experiments, where each dataset is a graph consisting of a set of vertices and a set of undirected edges. Each vertex represents a point (i.e., an object) in the road network, and each edge represents the road segment directly connecting two vertices. Since these datasets contain noise such as self-loop edges and unconnected graph segments~\cite{yao18}, we performed data pre-processing to remove them.
To quickly obtain the shortest-path distance $D$ between two objects (vertices), we used the original C/C++ source code written by the creators of the PHL algorithm\footnote{https://github.com/kawatea/pruned-highway-labeling}. Table~\ref{tab42} summarizes the parameters to be considered in the experiments, where the default parameter values are given in parentheses.

\begin{table}
\caption{Road network datasets.} \label{tab41}
\centering \tablefontsize
\begin{tabular}{cccc} \hline
Acronym & Name & Vertices & Edges \\ \hline
NY & New York City & 264,346 & 733,846 \\
COL & Colorado	& 435,666 & 1,057,066 \\
NW & Northwest USA & 1,207,945 & 2,840,208 \\
LKS & Great Lakes & 2,758,119 & 6,885,658 \\
W & Western USA & 6,262,104 & 15,248,146 \\ \hline
\end{tabular}
\end{table}

\begin{table}
\caption{Experiment parameters.} \label{tab42}
\centering \tablefontsize
\begin{tabular}{ccc} \hline
Parameter & Description & Values (default value) \\ \hline
$\R$ & road network dataset & NY, COL, NW, LKS, W (NW) \\
$M$ & size of $Q$, i.e., $|Q|$ & 64, 128, 256, 512, 1024 (256) \\
$k$ & number of nearest neighbors & 1, 5, 10, 15, 20 (1) \\
$\phi$ & flexibility factor & 0.1, 0.3, 0.5, 0.8, 1.0 (0.5) \\
$C$ & coverage ratio of $Q$ & 0.01, 0.05, 0.10, 0.15, 0.20 (0.10) \\ \hline
\end{tabular}
\end{table}

\begin{figure}[t]
\centering \capfontsize
\begin{tabular}{ccc}
\includegraphics[scale=\expscale]{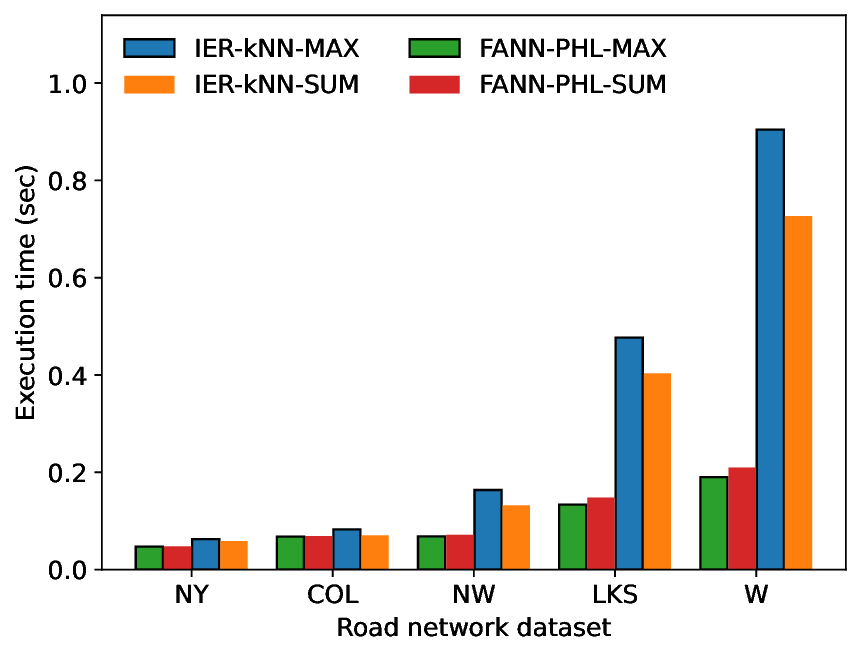} &
\includegraphics[scale=\expscale]{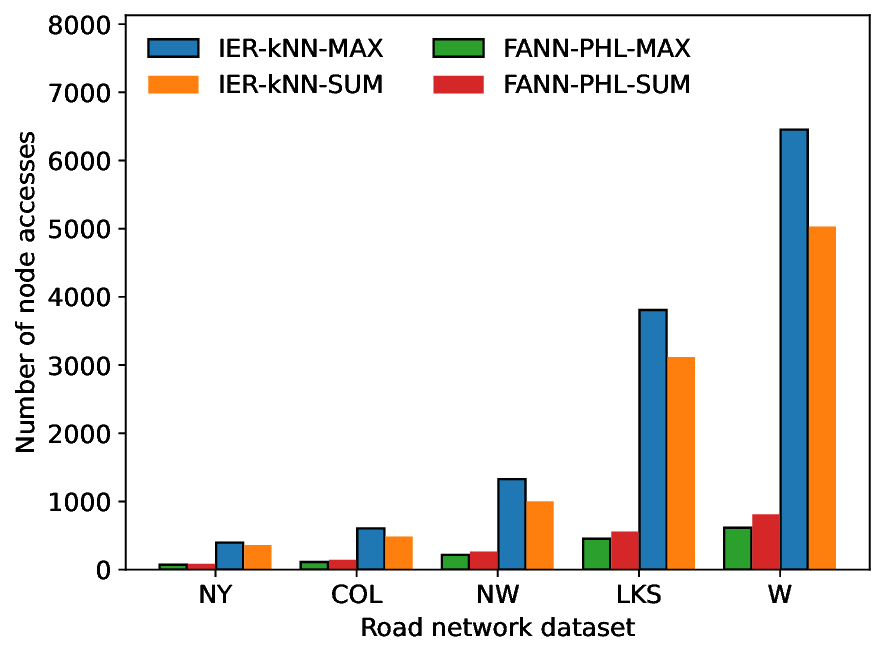}\\
(a) Execution time (seconds). & (b) Number of node accesses.
\end{tabular}
\caption{Comparison of FANN performance for various road network datasets ($\R$).} \label{exp-r}
\end{figure}

In the first experiment, we compared the execution time needed for FANN search and the number of index node accesses for all road network datasets listed in Table~\ref{tab41}. All the other parameters were set to the default values in Table~\ref{tab42}. Figure~\ref{exp-r} shows the results of the first experiment; the values in this figure are the averages of the results obtained by 1,000 randomly generated query sets. The results for the aggregate functions $\g$ = max and sum were represented by adding ``MAX'' and ``SUM'' to the names of the two algorithms, respectively, e.g., FANN-PHL-MAX and FANN-PHL-SUM. As shown in this figure, both FANN search algorithms showed similar trends in the execution time and the number of index node accesses. The number of objects included in the query region of the same size increased with the size of the road network. Therefore, the number of distance calculations to them and the execution time also increased. In the first experiment, FANN-PHL consistently outperformed IER-$k$NN with the improvement ratio of up to 4.75 times for the W dataset and $\g$ = max.

\begin{figure}[t]
\centering \capfontsize
\begin{tabular}{ccc}
\includegraphics[scale=\expscale]{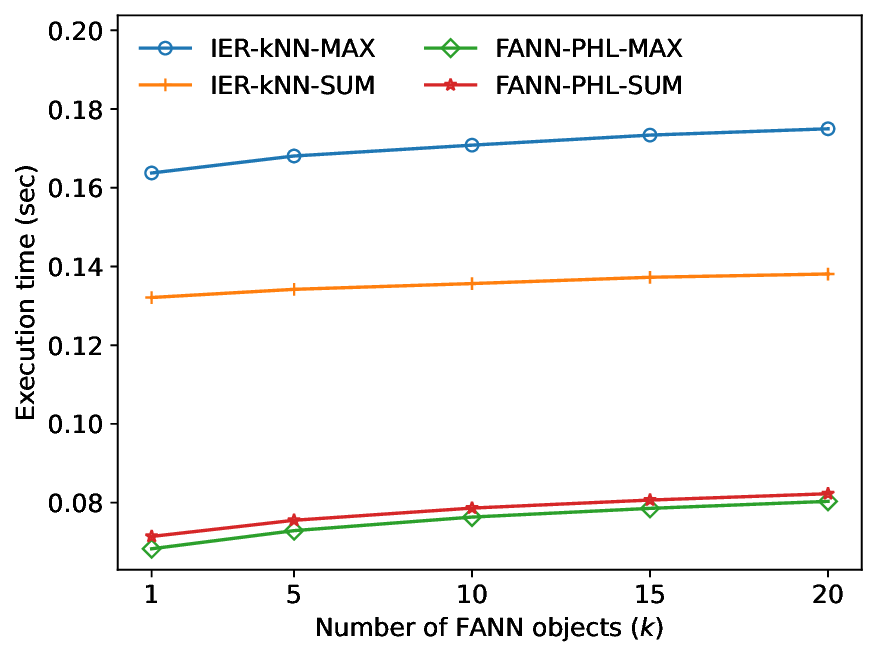} &
\includegraphics[scale=\expscale]{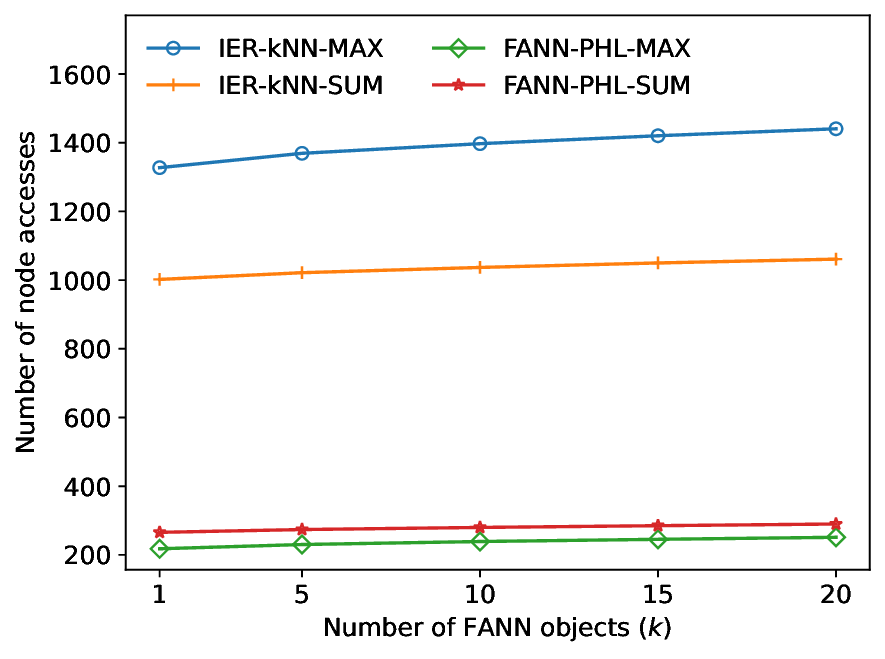}\\
(a) Execution time (seconds). & (b) Number of node accesses.
\end{tabular}
\caption{Comparison of FANN performance for various numbers of nearest neighbors ($k$).} \label{exp-k}
\end{figure}

In the second experiment, we compared the FANN search performance while changing the number of the nearest objects $k$, and the results are shown in Figure~\ref{exp-k}. For both FANN-PHL and IER-$k$NN, since the pruning bound increased with $k$, more index nodes were visited, and the execution time also increased. In this experiment as well, FANN-PHL consistently outperformed IER-$k$NN with a performance improvement of up to 2.40 times for $k$ = 1 and $\g$ = max.

\begin{figure}[t]
\centering \capfontsize
\begin{tabular}{ccc}
\includegraphics[scale=\expscale]{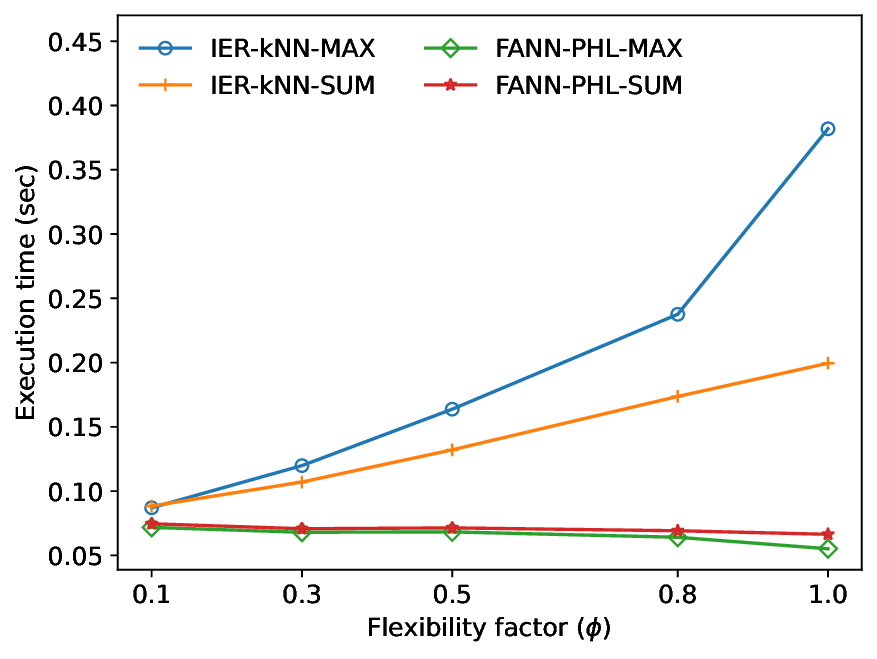} &
\includegraphics[scale=\expscale]{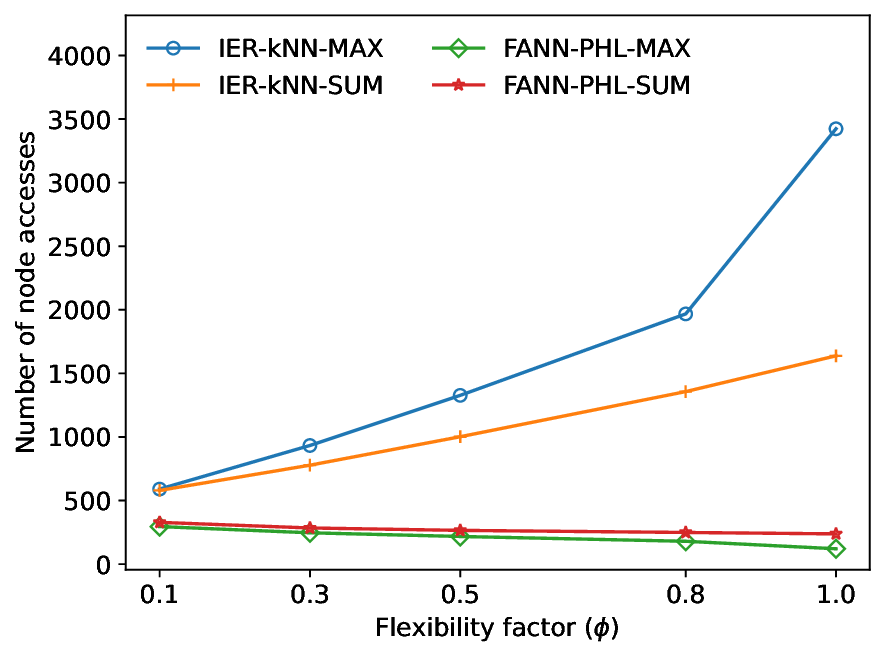}\\
(a) Execution time (seconds). & (b) Number of node accesses.
\end{tabular}
\caption{Comparison of FANN performance for various flexibility factors ($\phi$).} \label{exp-p}
\end{figure}

In the third experiment, we compared the performance of FANN search for various flexibility factor $\phi$ values, and the results are shown in Figure~\ref{exp-p}. It can be observed that, as $\phi$ increased, the execution time and the number of index node accesses of IER-$k$NN increased. This is because, for higher $\phi$, the FANN distance $\hat{p}^*.g_\phi$ of the FANN candidate object $\hat{p}^*$ becomes larger, and more R-tree nodes are visited. In contrast, for FANN-PHL, even with an increase in $\phi$, the execution time and the number of M-tree node accesses decreased. This is because, as $\phi$ increases in line~\ref{l1.08} in Algorithm~\ref{alg1}, $e'.g_\phi$ for an entry $e'$ also increases faster than $\hat{p}^*.g_\phi$.
When calculating $e'.g_\phi$, $Q_\phi$ is composed of the query objects closest to $e'$ among those in $Q$, so for a smaller $\phi$, it is likely that more query objects $q_i ~ (\in Q_\phi)$ are included in the spherical region of $e'$. Since we have $D(e', q_i) = 0$ for these $q_i$ as $q_2$ in Figure~\ref{fig32}, $e'.g_\phi$ also becomes zero or very close to zero. However, for a larger $\phi$, the probability decreases, and it is more likely that $e'.g_\phi > \hat{p}^*.g_\phi$.
Therefore, a smaller number of entries $e'$ are added into $H$ as $\phi$ increases. In this experiment as well, FANN-PHL consistently showed a better performance than IER-$k$NN with a performance improvement of up to 6.92 times for $\phi$ = 1.0 and $\g$ = max.

\begin{figure}[t]
\centering \capfontsize
\begin{tabular}{ccc}
\includegraphics[scale=\expscale]{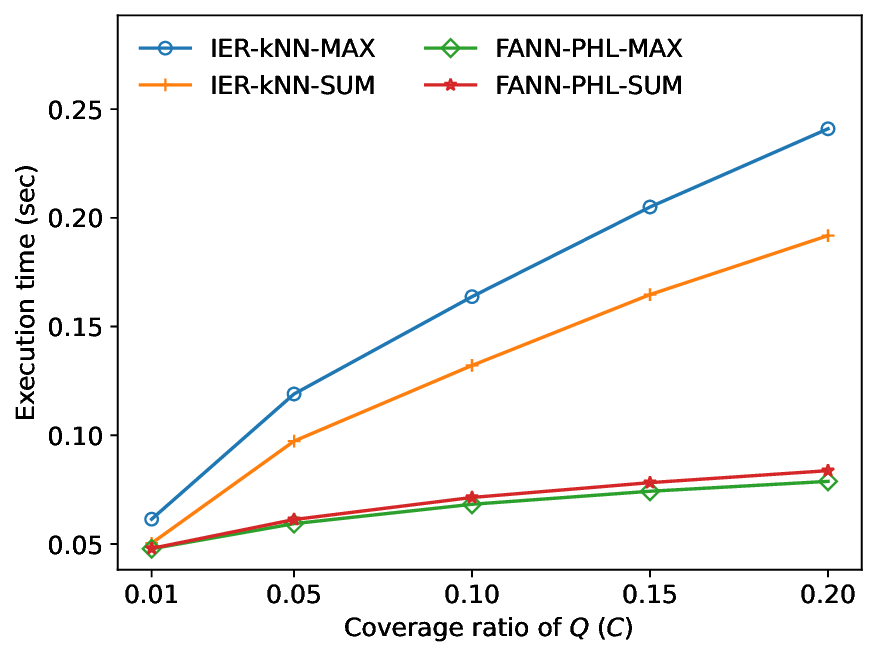} &
\includegraphics[scale=\expscale]{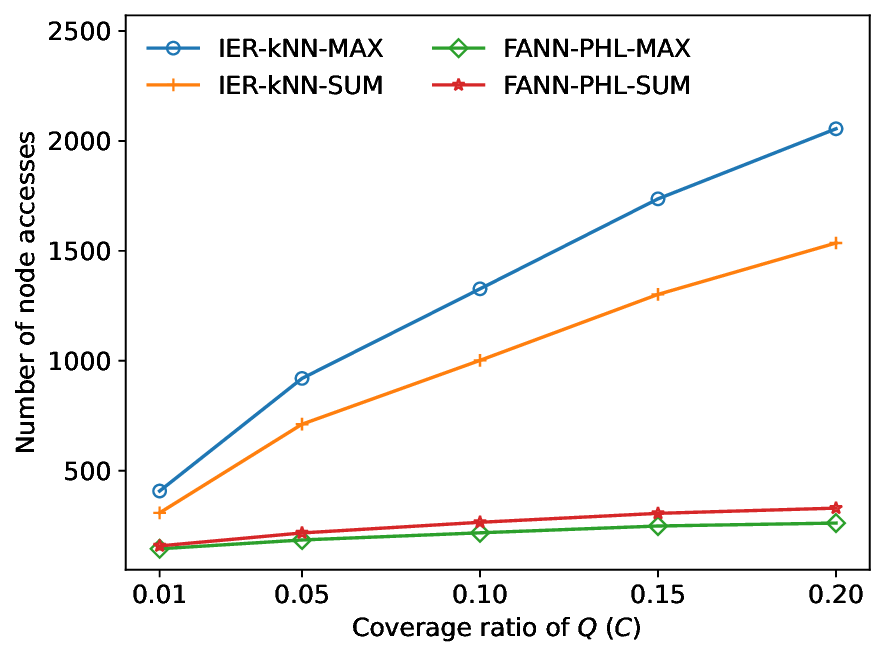}\\
(a) Execution time (seconds). & (b) Number of node accesses.
\end{tabular}
\caption{Comparison of FANN performance for various coverage ratios of query ($C$).} \label{exp-c}
\end{figure}

In the fourth experiment, we compared the performance of FANN search while changing the coverage ratio $C$ of query objects, where $C$ denotes the ratio of the minimum area including all query objects to the area occupied by all road network objects. Figure~\ref{exp-c} shows the experimental results. For higher $C$, the number of index nodes included in the query object area increases, and the execution time becomes larger. In this experiment, FANN-PHL consistently performed better than IER-$k$NN with a performance improvement of up to 3.06 times for $C$ = 0.2 and $\g$ = max.

\begin{figure}[t]
\centering \capfontsize
\begin{tabular}{ccc}
\includegraphics[scale=\expscale]{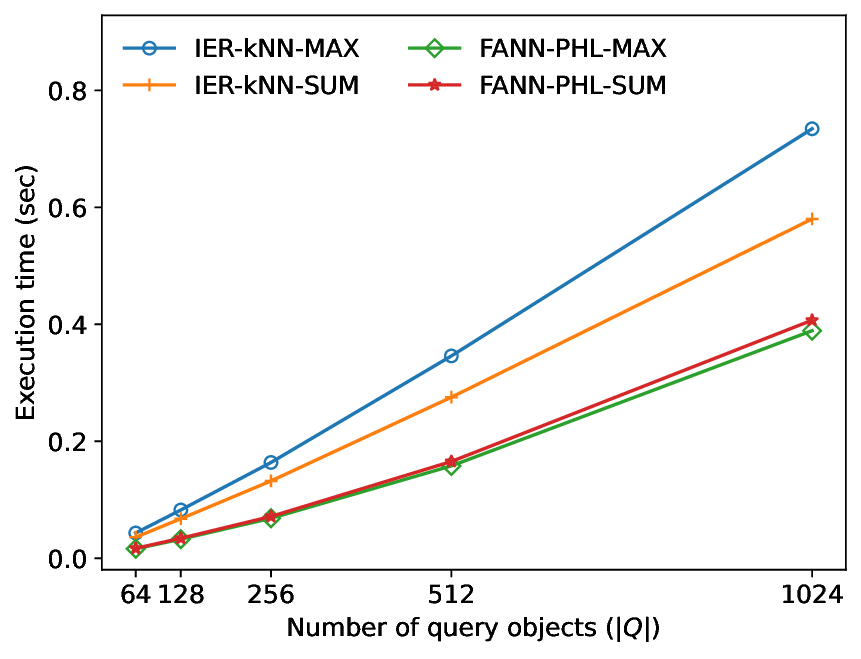} &
\includegraphics[scale=\expscale]{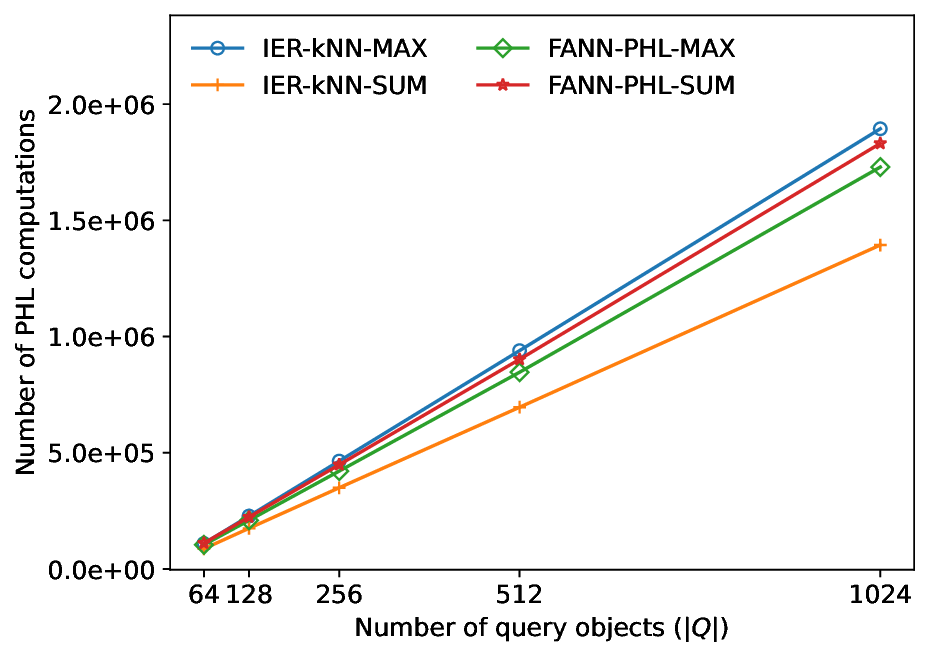}\\
(a) Execution time (seconds). & (b) Number of distance computations.
\end{tabular}
\caption{Comparison of FANN performance for various number of query objects ($M$).} \label{exp-m}
\end{figure}

In the final experiment, we compared the performance of FANN search while changing the number of query objects $M$, and the results are shown in Figure~\ref{exp-m}.
For both algorithms, we found that, as $M$ increased, the number of index node accesses remained almost constant while the execution time increased linearly. This is because, even though $M$ increases, there are no noticeable variations in $\hat{p}^*.g_\phi$ and $e'.g_\phi$ since the area of query objects remains similar. The number of M-tree nodes accessed by FANN-PHL was much smaller than the number of R-tree nodes accessed by IER-$k$NN. Meanwhile, as $M$ increased, the number of calculations of distance $D$ increased linearly for both algorithms as shown in Figure~\ref{exp-m}(b). This is because the actual distance $D$ to all $M$ query objects $q_i$ should be calculated to obtain $\hat{p}^*.g_\phi$ and $e'.g_\phi$. Owing to these two factors, the execution time of both algorithms increased linearly with $M$. In this experiment as well, FANN-PHL consistently outperformed IER-$k$NN with a performance improvement of up to 2.67 times for $M$ = 64 and $\g$ = max.

\section{Conclusions}
\label{sec:conclusions}

This study proposed the FANN-PHL algorithm for efficient exact $k$-FANN search using the M-tree~\cite{cia97}. The state-of-the-art IER-$k$NN algorithm~\cite{yao18} used the R-tree~\cite{man05} and pruned off the index nodes that are unlikely to include the final result object using the Euclidean distances. However, IER-$k$NN made many unnecessary accesses to index nodes and thus performed many calculations of the shortest-path distances to the objects included in the unnecessary nodes since the Euclidean distances are significantly different from the actual shortest-path distances between objects in road networks.
Our FANN-PHL algorithm can prune off the index nodes more accurately than IER-$k$NN by using the M-tree, which is constructed based on the actual distances between objects, and can also dramatically reduce the calculations of the shortest-path distances. To the best of our knowledge, FANN-PHL is the first exact $k$-FANN algorithm that uses the M-tree. We proved that our algorithm does not cause any false drop. Through a series of experiments using various real road network datasets, we demonstrated that FANN-PHL consistently outperformed IER-$k$NN for all datasets and parameters with a performance improvement of up to 6.92 times.

\section*{Acknowledgments}
\label{sec:ack}

This work was supported by the National Research Foundation of Korea (NRF) grant funded by the Korean government (MSIT) (No. 2021R1A2C1014432).
This work was also supported by the Institute of Information \& Communications Technology Planning \& Evaluation (IITP) Grant funded by the Korean government (MSIT) (No. 2020-0-00073, Development of Cloud-Edge-based City-Traffic Brain Technology).

\end{document}